\newcommand{\ud}{\rm d}
\newcommand{\un}{~\mathrm}

\documentclass[twocolumn,showpacs,aps,prl]{revtex4-1}

\usepackage{graphicx}
\usepackage{bm}
\usepackage{color}
\usepackage{epsfig}

\begin{document}

\preprint{APS/123-QED}

\title{Atomic-scale avalanche along a dislocation in a random alloy}

\author{S. Patinet$^1$, D. Bonamy$^2$ and L. Proville$^3$}
\affiliation{$^1$Laboratoire PMMH, UMR 7636 CNRS/ESPCI/Paris 6/Paris 7, 10, rue Vauquelin, 75231 Paris Cedex 5, France}
\affiliation{$^2$CEA, IRAMIS, SPCSI, Grp. Complex Systems and Fracture, F-91191 Gif sur Yvette, France}
\affiliation{$^3$CEA, DEN, Service de Recherches de M\'etallurgie Physique, F-91191 Gif-sur-Yvette, France}

\begin{abstract}
The propagation of dislocations in random crystals is evidenced to be
governed by atomic-scale avalanches whose the 
extension in space and the time intermittency
characterizingly diverge at the critical threshold.
Our work is the very first atomic-scale evidence that the paradigm of
second order phase transitions applies to the depinning of elastic interfaces in random media.
\end{abstract}


\maketitle

\section{INTRODUCTION}

Avalanche like motion of isolated elastic interfaces in random media is a process now clearly recognized at laboratory scales, e.g. in Barkhausen effect \cite{Durin2000}, brittle fracture \cite{MaloyPRL2006} imbibition \cite{santucci_avalanches_2011}, etc. 
Mesoscale scalar field theory \cite{NarayanFisherPRB1993} 
captures such a behavior and allows theoreticians to define an universal
critical behavior close to the depinning threshold, 
with a divergence of mean avalanche extents and 
durations. Though, the field theory remains limited to a
realm by far larger than the atomic-scale. This is due to
a prerequisited coarse-graining which 
introduces an averaging distance cutoff much larger than
the typical inter-atomic distances. 

Here we show that the critical avalanching behavior 
predicted throughout field theory is still a concern 
at the atomic-scale, along an isolated dislocation moving in a random crystal. Dislocations exibit morphological scaling features. They also propagate through jerky avalanches, the size and duration of which are power-law distributed up to a cut-off which diverges as the critical stress is approached. All the scaling relations expected from the standard depinning theory are fulfilled down to the atomic scale.

\section{NUMERICAL METHODS}

Plastic deformation of a solid solution is a prototypical example 
where dislocations must pass a random distribution of atomic-size obstacles 
to release plastic flow \cite{Friedel1964}. Despite the most recent progresses in tunneling electron microscopy \cite{Caillard2011}, it is practically unfeasible for experiments to analyze selectively a dislocation depinning and its associated roughness, at the atomic scale.
Hence numerical simulations are resorted to, as they enable us to focus on the dynamics of a  
single dislocation. Molecular dynamics (MD) simulations are employed in order to integrate the degrees of freedom of the whole crystal with atomic size impurities. The main advantage of MD simulations is that the dislocation is not modeled by a phenomenological elastic manifold  \cite{zapperidepinning2001,Csikor2007,Kubin_DDD} but more realistically as a Burgers discontinuity in the atom arrangement. Note also that this method ensures a proper inclusion of the non-linearities governing the dislocation self-interactions and a time scale with clear physical meaning.

\begin{figure}[!]
\includegraphics[width=0.9\columnwidth]{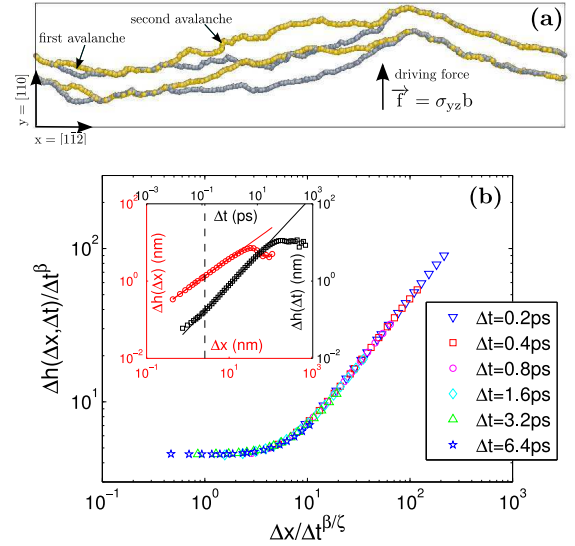}
\caption{
\label{fig1}(Color online) 
(a) Plane view for an edge dislocation at two different times
in Ni(Al) alloy, modeled within EAM Ref.\cite{Rodary2004}. Only Shockley partial dislocation core atoms are colored before (gray) and after (yellow) a sequence of two avalanches separated by $10$ ps. 
The rest of the crystal atoms are not shown. 
(b) Main panel: Collapsed space-time height-height
correlation function $\Delta h/\Delta t^{\beta}$ vs. 
$\Delta x/\Delta t^{\beta/\zeta}$ at various $\Delta t$ obtained at $\sigma_{yz}=171.2 \un{MPa}$ with $\zeta = 0.85$ and $\beta = 0.76$ obtained from the fits in inset.
Inset: space (circles) and time (squares) height-height
correlation function. Straight lines are power-law fits. The vertical dashed line corresponds to the lower time cutoff.}
\end{figure}
 
Typical snapshots of the dislocation position at two successive times are shown in Fig. \ref{fig1} (a). A single-crystal of binary Ni(Al) solid solutions (10 atomic percent of Al randomly distributed in the Ni matrix) with a single $[110]$ edge dislocation is deformed under a constant shear stress at $5$ K. The technical details of these MD simulations can be found in \cite{Rodary2004} and briefly recalled hereafter.

The interactions between atoms are modeled via an Embedded-Atom-Method interatomic potential which has been tailored to bulk properties \cite{Rodary2004}. Periodic boundary conditions are imposed in the dislocation line (X in Fig. \ref{fig1} (a)) and in the dislocation glide direction (Y in Fig. \ref{fig1} (a)). The size of the simulation box is $130\times10\un{nm}$ in the X and Y directions, respectively, and the thickness is $3.5\un{nm}$ [The size has been chosen to be several times larger than the Larkin's length]. The upper and lower free surfaces of the system (parallel to the slip) allow us to impose a shear 
stress $\sigma_{yz}$. This translates into a Peach-Koehler force that drags the dislocation
through the random crystal above a certain threshold studied in details in \cite{rodney_dislocation_1999,Rodary2004,Proville2010bis}.

The initial configuration is computed with a mere steepest descent gradient algorithm. Due to the relatively weak stacking fault energy of Ni, the dislocation dissociates into two Shockley partial dislocations as expected for face-centered cubic metals (see Fig. \ref{fig1} (a)). Following this initial relaxation, we apply a constant shear stress ($\sigma_{yz}$) and temperature ($5\un{K}$). The latter is maintained via a Berendsen thermostat. 

The dislocation dynamics are analyzed along the trajectories for which the glide distance  
is at least $150$ nm. With the present configuration, the closest stress from the critical depinning transition point is obtained at $\sigma_{yz}=171.2$ MPa. This order of magnitude is similar to that measured in experiments \cite{Mishima1986,Chan2001}. The dislocation core is localized by identifying the atoms whose first neighbor cell differs from the perfect crystal. This yields a limit $\ell_x=0.40$ nm (resp. $\ell_y=0.58$ nm) in resolution along X (resp. Y). 
The two partial dislocations move in a coherent manner (see Fig. \ref{fig1} (a)) due to their strong elastic coupling\cite{Proville2010bis}. As results, we define the effective dislocation position by averaging the two partials. All the analysis thereafter are restricted to the steady state regime, after the dislocation has glided over a distance of $20$ nm. Let us finally add that the high-frequency oscillations were removed from the dislocation dynamics and we only considered the forward motions in the statistical analysis presented hereafter.

\section{RESULTS}

{\em Morphological scaling features} of the time evolving fronts are
first characterized. In this context, the height-height correlation functions are computed in both space and time:

\begin{eqnarray}
\Delta h(\Delta x)&=\langle(h(x+\Delta x,t)-h(x,t))^{2}\rangle^{1/2},\nonumber\\
\Delta h(\Delta t)&=\langle(h(x,t+\Delta t)-h(x,t))^{2}\rangle^{1/2}
\end{eqnarray}

\noindent Both are found to exhibit power-law shapes, with exponents $\zeta = 0.85 \pm 0.05$ and $\beta =0.76 \pm 0.02$, respectively (Fig. \ref{fig1} (b): Inset). These scalings are signatures of self-affinity. No lower cutoff is evidenced in space, while the lower cutoff in time appears to be of the order of the inverse of the Debye frequency ($7.8\un{THz}$ in pure Ni). In both cases, the upper cutoffs decay as applied shear $\sigma_{yz}$ (and hence mean dislocation velocity) increases.
  
Full spatio-temporal morphological scaling features can be characterized through
the computation of the space-time structure function defined as:

\begin{equation}
\Delta h (\Delta x,\Delta t)=\langle(h(x+\Delta x,t+\Delta t)-h(x,t))^{2}\rangle^{1/2}
\end{equation} 

\noindent As shown in the main panel of Fig. \ref{fig1} (b), this function is found to obey 
the Family-Viseck scaling \cite{FamilyBOOK1991}:

\begin{equation}
\begin{array} {l}
\Delta h \propto \Delta t^{\beta}f(\frac{\Delta x}{\Delta t^{\beta/\zeta}}),~	f(u) \sim \left\{
\begin{array}{l l}
1 & $if$ \quad u \ll 1  \\
u^{\zeta} & $if$ \quad u \gg 1
\end{array}
\right\},
\end{array}
\label{cor2D}
\end{equation}

\noindent Such a scaling is expected close to the depinning transition of an elastic manifold. Hence, the exponents $\zeta$, $\beta$ and $z=\zeta/\beta=1.12 \pm 0.1$ are identified with the roughness, growth and dynamic exponents, respectively, commonly defined in interface growth problems.

\begin{figure}
\includegraphics[width= 0.9\columnwidth]{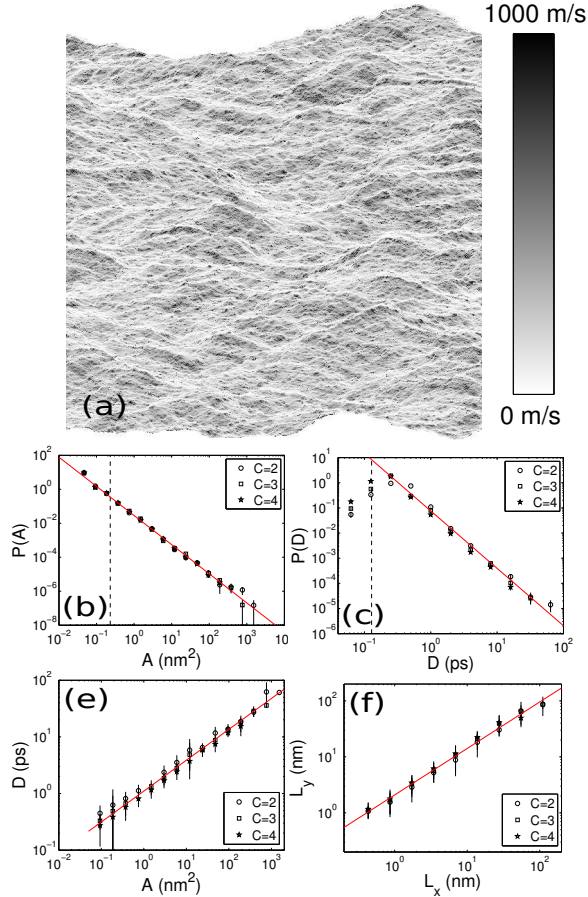}
\caption{(Color online) (a) Gray-scale map of velocity matrix obtained in 
our simulation at $\sigma_{yz}=171.2\un{MPa}$. (b) Area distribution of
avalanches $P(A)$ as defined in the text for different clip levels (see legend). (c) Duration distribution of avalanches $P(D)$ with same clip levels. 
Error bars are computed as in \cite{BergCPC2008}.
(d-e) The scaling of $D$ vs. $A$ and that of avalanche width $L_y$ vs. avalanche length $L_x$, respectively. In both plots, errorbars indicate the standard deviation. Full lines correspond to power-law fits: $P(A) \propto A^{-\tau}$, $P(D) \propto A^{-\alpha}$, 
$D\propto A^\gamma$ and  $L_y \propto L_x^H$ with $\tau=1.71\pm 0.03$, $\alpha=2.28\pm 0.1$, $\gamma=0.55\pm 0.03$ and $H =0.83 \pm 0.04$, respectively. The $\pm$ range is defined for a $95\%$ confident interval. The vertical dashed lines correspond to the resolution limits in space and time.}
\label{fig2}
\end{figure}

{\em Spatio-temporal intermittent dynamics} of the propagating 
dislocations are now analyzed via a procedure 
initially proposed in \cite{MaloyPRL2006} and extensively applied to 
crack propagation \cite{BonamyPRL2008,BonamyJPD2009} 
and imbibition \cite{santucci_avalanches_2011} problems, among others.
It consists: (i) in computing the so-called activity map $\bf{w}$, i.e. the time w$(x,y)$ spent by the dislocation within a small $2.16 \times 2.16$ \AA$^2$ region at each point $(x,y)$
of the glide plane (Fig. \ref{fig2} (a)); and (ii) in subsequently defining 
avalanches as clusters of connected points with velocity $v=1/$w above $v_c=C\langle v \rangle$ where $\langle \rangle$ denotes averaging over both time and space, and $C$  stands for clip level. The statistics of avalanche area, $A$, and duration, $D$, (time elapsed between dislocation arrival and departure in/from the cluster) allow us to characterize quantitatively the intermittent dynamics.

Dislocation propagation just above the depinning threshold ($\sigma_{yz} = 171.2\un{MPa}$) is first considered. Distributions of area $A$ and duration $D$ are presented in Fig.\ref{fig2} (b)-(c). They exhibit power law tails 
$P(A) \propto A^{-\tau}$ and $P(D) \propto D^{-\alpha}$, as expected in
a system near criticality. The two exponents are found to be 
$\tau = 1.71\pm 0.03$ and $\alpha = 2.28 \pm 0.1$. Figure \ref{fig2} (c)
reveals also a power-law scaling between area and duration: $D \propto A^{\gamma}$
 with $\gamma = 0.55 \pm 0.03$. Note that $\tau$, $\alpha$ and $\gamma$ are related:
Since $D$ scales as $A^{\gamma}$, the fact that $P(A) \propto D^{-\tau}$ yields $P(D) \propto A^{-(1+(\tau-1)/\gamma)}$ and therefore: $\alpha=1+(\tau-1)/\gamma$. The fulfillment of this relation
indicates the high quality of our sampling. Finally, 
avalanches are shown to exhibit morphological scaling features since their width, $L_y$, scales as a power-law of their length, $L_x$ (Fig. \ref{fig2} (e)), with an exponent $H=0.83 \pm 0.03$. All 
the scaling exponents described above are found to be robust and independent of clip level.

\begin{figure}
\includegraphics[width= 0.9\columnwidth]{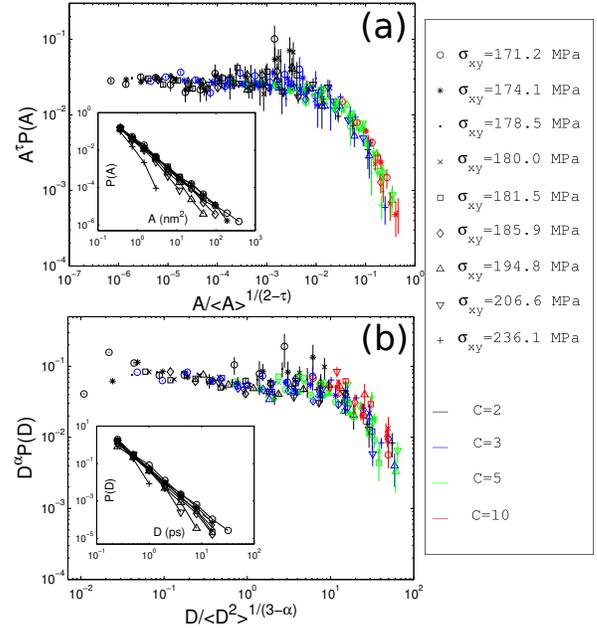}
\caption{(Color online) 
(a) Main panel: Collapsed area distribution obtained from Eq. \ref{distA} with $\tau=1.71$ fitted from data in Fig. \ref{fig2} (b). Symbols correpond to different clip levels, $C=2$ (black), $C=3$ (blue), $C=5$ (green) and $C=10$ (red) and various $\sigma_{yz}$, $171.2$ (circle), $174.1$ (asterisk), $178.5$ (point), $180$ (cross), $181.5$ (square), $185.9$ (diamond), $194,8$ (up triangle), $206,6$ (down triangle) and $236,1$ MPa (plus) according to the right-handed legend. Inset: Avalanche size distribution at $C=3$ for the different $\sigma_{yz}$ used in main panel (same symbols). (b) Main panel: Collapsed duration distribution obtained using Eq. \ref{distD} with $\alpha= 2.28$ fitted from Fig. \ref{fig2} (c). Inset: Avalanche duration distribution for fixed clip level $C=3$ and various $\sigma_{yz}$ (same values as in (a)).
}
\label{fig3}
\end{figure}

Let us now move to the analysis of the role played by the applied stress $\sigma_{yz}$
on the dislocation dynamics. The insets in Figs. \ref{fig3} (a) and (b) show the
effect of applied stress on the area and duration distributions,
respectively. In both cases, an upper cutoff decreasing with $\sigma_{yz}$ is observed. Once again, such behaviors are characteristic
of a system near criticality. In this context, the distributions are expected to take the form
$P(A)=A^{-\tau} f(A/A_0)$ and $P(D)=D^{-\alpha} g(D/D_0)$ where $f$ and $g$ are universal
fast decreasing functions, and $A_0$ and $D_0$ are the area and duration cutoffs that diverge 
algebraically as $\sigma_{xy}$ reaches the critical value. Then, 
the mean value $\langle A \rangle=\int_0^\infty A \times P(A) \ud A$ goes as 
$A_0^{2-\tau}$, and the area distribution can be recast:
\begin{equation}
P(A) = A^{-\tau} f (A/\langle A \rangle^{1/(2-\tau)}).
\label{distA}
\end{equation}
As shown in the main panel in Fig. \ref{fig3} (a), this scaling is fully verified, 
and the function $f$ is found to be independent of both $\sigma_{yz}$ and $C$ over the whole range
tested. Note that this analysis must be adapted for durations since $\langle D \rangle=\int_0^\infty D \times P(D) \ud D$ is not defined when $\alpha$ is larger than two. Instead, the mean value $\langle D^2 \rangle$ is chosen. Since $\langle D^2 \rangle$ goes as $D_0^{3-\alpha}$, one hence expects: 
\begin{equation}
P(D) = D^{-\alpha} g (D/\langle D^2 \rangle^{1/(3-\alpha)}).
\label{distD}
\end{equation}
This second scaling is fairly well fulfilled, as demonstrates Fig. \ref{fig3} (b). 

\begin{figure}
\begin{center}
\includegraphics[width=0.99\columnwidth]{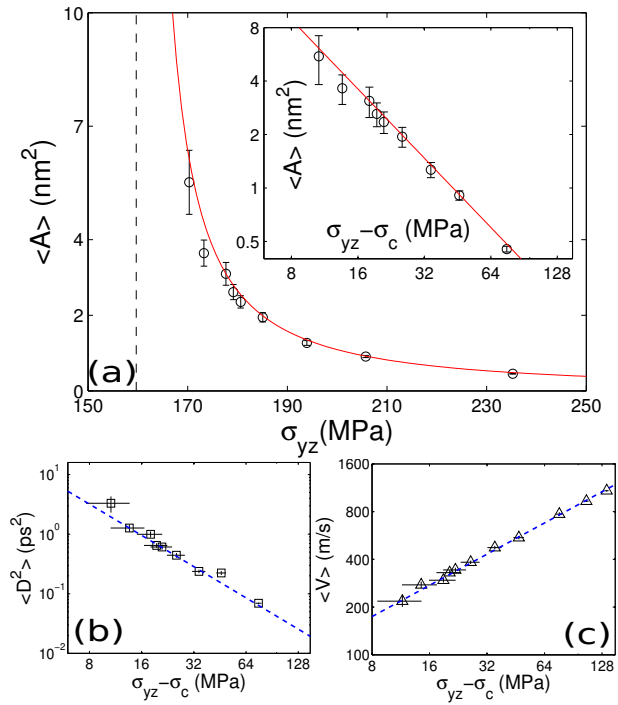}
\caption{(Color online) (a) Main panel: Variation of the mean avalanche area $\langle A \rangle$ 
with applied stress $\sigma_{yz}$. Full line is an algebraic divergence 
$\langle A \rangle \propto (\sigma_{yz} - \sigma_c)^{-\nu_A}$ with fitted
 parameters $\sigma_c = 159.7 \pm 3\un{MPa}$ and 
$\nu_A = 1.29 \pm 0.05$. Vertical dashed line shows the 
position of $\sigma_c$. Inset: Log-log plot of $\langle A \rangle$ 
vs $\sigma_{yz} - \sigma_c$. The straight line corresponds to the fit. (b) Log-log plot of the mean value of square duration $\langle D^2 \rangle$ vs. $\sigma_{xy} - \sigma_c$. The straight dashed line is obtained with $\nu_D = 1.75$ from scaling arguments (not fitted). (c) Same as (b) for mean velocity with exponent $\theta = 0.65$ (not fitted).}
\label{fig4}
\end{center}
\end{figure}

To complete the statistical analysis of dislocation dynamics, we plot 
in Figs. \ref{fig4} (a) the variation of $\langle A \rangle$ as a function of
$\sigma_{yz}$. Once again, the increase of $\langle A \rangle$ as $\sigma_{yz}$
decreases toward the depinning threshold is reminiscent of a critical behaviour. It is well described by an algebraic divergence of the form 
$\langle A \rangle \propto (\sigma_{yz} - \sigma_{c})^{-\nu_A}$ with $ \nu_A = 1.29 \pm 0.05$. This divergence in terms of avalanche
area translates into a divergence in term of avalanche duration: 
Since $\langle D^2 \rangle \propto D_0^{(3-\alpha)}$ and $D \propto A^\gamma$, one expects
$\langle D^2 \rangle \propto A_0^{\gamma(3-\alpha)} \propto \langle A \rangle^{\gamma(3-\alpha)/(2-\tau)}$ 
and hence, $\langle D^2 \rangle \propto (\sigma_{yz} - \sigma_{c})^{-\nu_D}$ with
$\nu_D = \nu_A\gamma(3-\alpha)/(2-\tau) = 1.75 \pm 0.34$. As seen in Fig. \ref{fig4} (b), 
this scaling is compatible with direct simulations. 

\section{CONCLUDING DISCUSSION}

Our analysis of  atomic-scale dislocation dynamics in random crystals has revealed many signatures of criticality. To the very best of our knowledge, this work is the first to show that the theoretical framework of second order phase transitions can be applied down to the atomic scales. 
In this respect, it is of some interest to check the theoretical mapping between the avalanche dynamics extracted from the activity map on one hand, and the revealed morphological scaling features on the other hand \cite{NarayanFisherPRB1993}: Avalanches of extent $L_x$ are indeed expected 
to result from front pieces of extent $L_x$ that depins over a propagating length $L_x^\zeta$
and a time $L_x^{1/z}$. Avalanche exponents $H$ and $\gamma$ are hence expected
to be related to $\zeta$ and $z$ through $H=\zeta$ and $\gamma = z/(1+\zeta)$ -- Both relations
are fulfilled in the present atomic scale system.

In the paradigm of critical transitions, the divergence of $\langle A \rangle$ and $\langle D^2 \rangle$
as $\sigma_{yz}- \sigma_c$ vanishes are due to the divergence of a correlation length $\xi$: 
$\xi \propto (\sigma_{yz}- \sigma_c)^{-\nu}$. Assimilating $\xi$ to the maximum extent of an avalanche
yields $\xi \approx A_0^{1/(1+\zeta)} \propto \langle A \rangle^{1/(1+\zeta)(2-\tau)}$ and hence 
$\nu = \nu_A/(1+\zeta)(2-\tau) = 2.44\pm 0.29$. Scaling of the mean dislocation velocity $\langle v \rangle$ can then be deduced: close to $\sigma_c$, the propagation is made of avalanches of extent
$\xi$, which moves the dislocation forward by $\xi^\zeta$ over a time period $\xi^z$. Thus, the
velocity behaves as $\langle v \rangle \approx \xi^\zeta/\xi^z \propto (\sigma_{xy}- \sigma_c)^{\theta}$ with $\theta=\nu(z-\zeta)$. This latter relation yields $\theta=0.65\pm0.3$, which is perfectly compatible with the direct measurements (Fig. \ref{fig4} (c)).

Our system exhibits all signatures expected from the field theory in the vicinity of a critical depinning transition at zero temperature. We can reasonably expect that the predictions of such a theory also apply at finite temperature. In particular, the slow thermally-activated motion of dislocation at low shear stress (below critical threshold) is expected to be given by the creep formula $<v> \propto \exp({-U_{0}(\sigma_{c}/\sigma_{xy})^{\mu}/kT})$ where $U_{0}$ is a characteristic energy scale and $\mu$ a universal exponent \cite{Feigelman1989,Nattermann1990}. Real deformation experiments actually result from a collective behavior of dislocation involving diverse mechanisms as the interactions with grain boundaries, surfaces, forest dislocations and atomic scale impurities. It seems therefore challenging to demonstrate this in real experiment and atomistic simulation could be a solution.

The present study convincingly shows that the concepts of critical depinning transition apply down to atomic scale. It is worth to mention that the critical exponents measured here do not belong to the standard universality classes associated with the well-established field theories of depinning transition, namely the Edward-Wilkinson (EW)\cite{Edwards1982}, the Kardar-Parisi-Zhang (KPZ)\cite{Kardar1986} and the Long Range (LR)\cite{Ertas1994} elastic string models (see Table \ref{Tab2}). This can be understood since these continuous string models are constructed from symmetries principles by calling upon a {\em thermodynamic limit} argument (i.e. by making $t \rightarrow \infty$ and $x \rightarrow  \infty$ to eliminate the high order derivates in the equation of string motion), which stops to be relevant at small scales. To understand what determines the universality class at such atomic scales represents a significant challenge for future investigations. Since the field theory of elastic manifold predicts critical behavior in variety of different systems of solid state physics like e.g. domain wall motion in ferromagnets, crack problems, vortex motion in superconductors and charge density wave, etc, it may be of interest to see whether or not similar extension of criticality down to atomic scale can be evidenced in these systems.

\begin{table*}
\caption{Depinning exponents obtained directly from simulations and from scaling relations (marked with ${*}$)}
\label{Tab}
\begin{tabular}{c|c|c|c}
$\zeta=0.85\pm0.05$ & $\beta=0.76\pm0.02$ & $z=1.12\pm0.1$ & $\theta_{*}=0.65\pm0.3$ \\
$\nu_{*}=2.44\pm0.29$ & $\tau=1.71\pm0.03$ & $\alpha=2.28\pm0.1$ & $H=0.83\pm0.04$    \\
$\gamma=0.55\pm0.03$ & $\nu_A=1.29\pm0.05$ & $\nu_{D*}=1.75\pm0.34$ &\\ 
\end{tabular}
\end{table*}

\begin{table*}
\caption{Comparison of the exponents measured in the atomic scale MD simulations reported here with those observed in some well-established scalar field models of elastic line depinning.}
\label{Tab2}
\begin{tabular}{|l|c|c|c|c|c|}
\hline
\hline
Present study & $\zeta=0.85\pm0.05$ & $\beta=0.76\pm0.02$ & $z=1.12\pm0.1$ & $\nu=2.44\pm0.29$ & $\theta=0.65\pm0.3$ \\
\hline
EW \cite{Duemmer2005} & $\zeta\simeq 1.26$ & $\beta\simeq 0.84$ & $z\simeq 1.5$ & $\nu \simeq 1.29$ & $\theta\simeq 0.33$ \\
KPZ \cite{Tang1992} & $\zeta\simeq 0.633$ & $\beta \simeq 0.633$ & $z\simeq1$ & $\nu\simeq 1.733$ & $\theta\simeq 0.636$ \\
LR \cite{Duemmer2007}& $\zeta\simeq 0.385$ & $\beta\simeq 0.5$ & $z\simeq 0.77$ & $\nu\simeq 1.625$ &  $\theta\simeq 0.625$ \\
\hline
\hline
\end{tabular}
\end{table*}


%

\end{document}